# Lack of superconductivity in the phase diagram of single-crystalline Eu(Fe$_{1-x}$Co$_x$)$_2$As$_2$ grown by transition metal arsenide flux

Gang Wang,[1,2,*] William R. Meier,[1,3] Warren E. Straszheim,[4] Joshua Slagle,[1] Sergey L. Bud'ko,[1,3] and Paul C. Canfield[1,3]

[1]Ames Laboratory, Iowa State University, Ames, Iowa 50011, USA

[2]Research and Development Center for Functional Crystals, Beijing National Laboratory for Condensed Matter Physics, Institute of Physics, Chinese Academy of Sciences, Beijing 100190, China

[3]Department of Physics and Astronomy, Iowa State University, Ames, Iowa 50011,

USA

[4]Civil and Construction Engineering Department, Iowa State University, Ames, Iowa

50011, USA

The interplay of magnetism and superconductivity (SC) has been a focus of interest in condensed matter physics for decades. EuFe$_2$As$_2$ has been identified as a potential platform to investigate interactions between structural, magnetic, electronic effects as well as coexistence of magnetism and SC with similar transition temperatures. However, there are obvious inconsistencies in the reported phase diagrams of Eu(Fe$_{1-x}$Co$_x$)$_2$As$_2$ crystals grown by different methods. For transition metal arsenide (TMA)-flux-grown crystals, even the existence of SC is open for dispute. Here we re-examine the phase diagram of single-crystalline Eu(Fe$_{1-x}$Co$_x$)$_2$As$_2$ grown by TMA flux. We found that the lattice parameter *c* shrinks linearly with Co doping, almost twice as fast as that of the tin-flux-grown crystals. With Co doping, the spin-density-wave (SDW) order of Fe sublattice is quickly suppressed, being detected only up to x = 0.08. The magnetic ordering temperature of the $Eu^{2+}$ sublattice ($T_{Eu}$) shows a systematic evolution with Co doping, first going down and reaching a minimum at x = 0.08, then increasing continuously up to x = 0.24. Over the whole composition range investigated, no signature of SC is observed.

## I. INTRODUCTION

The interplay of magnetism and superconductivity (SC) has been a reoccurring theme in condensed matter physics for several decades [1,2]. It has been indicated by more and more evidence that unconventional SC, such as in heavy fermion, cuprates, and iron-based superconductors, is closely related to magnetism [3,4]. $EuFe_2As_2$ (Eu-122) is a parent member of iron-based superconductors containing a local-moment *4f* atom. It develops not only spin-density-wave (SDW) order of the Fe sublattice below 200 K, which is typical for 122-based iron pnictides, but also an additional antiferromagnetic order of $Eu^{2+}$ moments below 20 K [5–7]. Meanwhile, there are reports that SC can be induced by suppressing the SDW order using chemical substitution or pressure [8–21]. It therefore potentially provides a good platform to investigate the interplay of magnetism and SC.

For Co-doped Eu-122, a superconducting transition at 21 K was first claimed in $Eu(Fe_{0.89}Co_{0.11})_2As_2$ grown by transition metal arsenide (TMA) flux, even though zero resistance was not achieved [10]. Similar phenomena were also observed by Ying et al. and Chen et al. [22,23]. The disturbing absence of zero resistance and re-entrance of resistance were attributed to the competition between SC and $Eu^{2+}$ magnetic order. Whereas zero resistance has been observed in $Sr_{1-y}Eu_y(Fe_{0.88}Co_{0.12})_2As_2$ crystals grown by TMA flux, there was no claim of coexisting SC and Eu-based ferromagnetism (FM) [24]. Up to now, zero resistance in $Eu(Fe_{1-x}Co_x)_2As_2$ has been realized only in the crystals with $x$ ranging from 0.15 to 0.20 grown by tin flux, where the superconducting onset temperature ($T_c^{onset}$) varies from 7.5 K to 20.5 K [11,25–28]. For tin-flux-grown $Eu(Fe_{1-x}Co_x)_2As_2$ crystals with $0.10 \leq x \leq 0.18$, it was claimed that strong FM from the Eu sublattice coexists with SC [28].

Given that SC is only seen with zero resistance for a rather narrow composition range in tin-flux-grown crystals, it is important to recall that comparison of TMA and tin-flux-grown Co-doped $CaFe_2As_2$ crystals clearly demonstrated that tin-flux-grown samples often have problems associated with homogeneity and reproducibility of Co-doping levels [29]. Therefore, we decided to determine the phase diagram of single-crystalline $Eu(Fe_{1-x}Co_x)_2As_2$ grown by TMA flux and carefully investigate how the magnetic order of $Eu^{2+}$ moments develops with Co doping and whether there is, indeed, any SC for this magnetism to interact with at all. We found that the growth technique influences both the lattice parameters and physical properties of $Eu(Fe_{1-x}Co_x)_2As_2$ crystals. No SC has been observed in our TMA-flux-grown $Eu(Fe_{1-x}Co_x)_2As_2$ single crystal with $x$ up to 0.24, which means, unlike tin-flux-grown crystals, there is no coexistence of SC and magnetism. A phase diagram of

TMA-flux-grown Eu(Fe$_{1-x}$Co$_x$)$_2$As$_2$ single crystals is established based on these results.

## II. EXPERIMENT

In this work, Eu(Fe$_{1-x}$Co$_x$)$_2$As$_2$ single crystals with nominal $x = 0$ (denoted as Co-0), 0.04 (Co-0.04), 0.08 (Co-0.08), 0.10 (Co-0.10), 0.12 (Co-0.12), 0.18 (Co-0.18), and 0.24 (Co-0.24) were grown by high temperature solution method out of TMA flux [30]. Given that the nominal and experimentally determined $x$-values are essentially the same (See Table I below) we will use the nominal $x$-values to identify the samples. The Fe$_{0.512}$As$_{0.488}$ and Co$_{0.512}$As$_{0.488}$ precursors were synthesized from ground As lumps (Alfa Aesar 99.9999%) and Fe powder (Alfa Aesar 99.9+%) or Co powder (Alfa Aesar 99.9+%) in a 1 : 1.05 atomic ratio in an argon filled fused-silica ampoule by solid-state reaction [31]. Eu rod (Material Preparation Center (MPC), Ames Laboratory 99.99%) was combined with ground, pre-reacted Fe$_{0.512}$As$_{0.488}$ and Co$_{0.512}$As$_{0.488}$ precursors in a ratio of Eu : Fe$_{0.512}$As$_{0.488}$ : Co$_{0.512}$As$_{0.488}$ = 1 : 6 (1 - $x$) : 6$x$ in a fritted, alumina crucible set (Canfield Crucible Set or CCS) [32]. The CCS filled with materials of a total mass of roughly 2 grams was sealed in a fused-silica ampoule filled with roughly 0.2 atmosphere argon. The ampoule was put into the muffle furnace and heated over 1 hour to 650 °C, held for 3 hours, then heated over 2 hours to 1150 °C, held at the temperature for 5 hours, cooled to 1050 °C over 0.5 hour, and then slowly cooled to 980 °C over 36 hours. At 980 °C, the assembly was removed from the furnace, inverted, and centrifuged to separate the remaining liquid from crystal. Single-crystalline Eu(Fe$_{1-x}$Co$_x$)$_2$As$_2$ grows as metallic plates up to several mm wide and several hundred microns thick (see Fig. 1). The crystals do not appear to be air sensitive and have no noticeable degradation of appearance and properties in air for several months.

X-ray diffraction data (XRD) were collected at room temperature on a Rigaku MiniFlex II powder diffractometer with Cu-$K_\alpha$ radiation and a graphite monochromator. Elemental analyses were carried out using a Scanning Electron Microscope (SEM, JEOL JXA-8200) equipped with wavelength-dispersive spectrometers (WDS). For each Eu(Fe$_{1-x}$Co$_x$)$_2$As$_2$ crystal, 12 spots were measured in different areas in order to better characterize the chemical composition over a wide area of the sample. Temperature and field-dependent magnetization and resistance measurements were carried out on a Quantum Design (QD) Magnetic Property Measurement System (MPMS). Samples for XRD, magnetization, and resistance measurements were cleaved along (*00l*) from thicker Eu(Fe$_{1-x}$Co$_x$)$_2$As$_2$ crystals using a razor blade. Susceptibility was measured with applied field parallel to the *ab* plane. The in-plane ac (f = 16Hz, I = 1 mA) resistance measurements were performed in QD MPMS systems operated in External Device Control (EDC) mode in conjunction with Linear Research LR700 ac resistance bridges. The samples for resistance measurements were cut into bars of (1-2.5) × (0.7-1.2) × (0.02-0.09) mm$^3$. Contacts for standard four-probe configuration were made by attaching platinum wires using silver epoxy, resulting in a contact resistance less than 5 Ω.

## III. RESULTS AND DISCUSSION

Figure 2a shows the XRD patterns for Eu(Fe$_{1-x}$Co$_x$)$_2$As$_2$ crystals. Only the (*00l*)

diffraction peaks with even $l$ are observed, indicating that the crystallographic $c$ axis is perpendicular to the plate surface [33]. As shown by the enlarged (*008*) diffraction peak (Fig. 2b), the peak position systemically shifts to higher angle with Co doping, which means that the lattice parameter $c$ shrinks. WDS analyses in Table I indicate that the real Co content is quite near to the nominal Co content and increases linearly with the nominal one. The composition variation from spot to spot is small, which suggests that the Co concentration is essentially homogeneous. The indexing results of XRD patterns for Eu(Fe$_{1-x}$Co$_x$)$_2$As$_2$ crystals in Fig. 2(a) further confirm the nearly linear shrinkage of lattice parameter $c$ with Co doping. Comparing the lattice parameter $c$ in Fig. 3 for crystals grown by TMA flux (this work) and tin flux [25,26], we found that they develop with Co doping in very different ways. The lattice parameter $c$ for the TMA-flux-grown crystals shrinks much faster upon Co doping than that of the tin-flux-grown crystals.

Figure 4a shows the temperature-dependent dc magnetic susceptibility of Eu(Fe$_{1-x}$Co$_x$)$_2$As$_2$ single crystals below 25 K measured under a small applied field of 10 Oe parallel to the *ab* plane in the zero-field-cooling (ZFC) protocol. The magnetic transition temperatures are determined by minimum $d(\chi T)/dT$ as shown in Fig. 4b. The first thing to note is that there is no indication at all of a superconducting phase transition. With Co doping, the ordering temperature of the Eu$^{2+}$ moments ($T_{Eu}$) decreases from 18.2 K for Co-0, through 17.4 K for Co-0.04, to 16.6 K for Co-0.08. Then $T_{Eu}$ increases continuously from 16.6 K to 18.4 K for Co-0.24. It looks like that the ordering of Eu$^{2+}$ first gets slightly suppressed with Co doping and then is strengthened after reaching a minimum value around $x = 0.08$. It is noticeable that there is another magnetic feature emerging at lower temperatures for Eu(Fe$_{1-x}$Co$_x$)$_2$As$_2$ single crystals with $x \geq 0.10$. The transition temperature ($T^*$) increases and is gradually drawn to $T_{Eu}$ with Co doping. Further investigations, using more techniques, will be needed to understand the origin of this feature.

The isothermal magnetization data for the Eu(Fe$_{1-x}$Co$_x$)$_2$As$_2$ single crystals are shown in Fig. 5. For Co-0, $M_{ab}$ initially increases almost linearly with the applied magnetic field, whereupon increases more rapidly and then above 0.8 T, gradually saturates to 7.56 $\mu_B$ near 5 T. The saturated value is a little bit larger than the theoretical saturated magnetic moment 7 $\mu_B$ of Eu$^{2+}$, which could be due to the contribution of Fe$^{2+}$ moments. Similar increasing trend of $M_{ab}$ could be seen for Co-0.04, and is then totally suppressed by Co doping.

Figure 6a shows the temperature-dependent normalized in-plane resistance ($R_{ab}$) for the Eu(Fe$_{1-x}$Co$_x$)$_2$As$_2$ single crystals. The temperatures for SDW order and the ordering of Eu$^{2+}$ moments are determined by $dR/dT$ as shown in Fig. 6b. The first, *exceptionally clear* observation is that there is no hint of SC in any of these samples. Given that even filamentary SC can give rise to conspicuous features in temperature dependent resistivity, these data further confirm that for Eu(Fe$_{1-x}$Co$_x$)$_2$As$_2$ crystals grown out of TMAs flux there is no conspicuous superconducting phase spanning any of the compositions we studied. For Co-0, the resistance shows a sharp transition at 178.9 K, associated with the SDW order and the structural transition [35]. With Co doping, the SDW order is quickly suppressed to 125.8 K for Co-0.04, 52.5 K for

Co-0.08, and then is not detectable for $x \geq 0.10$. The kink below 20 K is due to the ordering of $Eu^{2+}$ moments, the temperature of which initially decreases and then increases, follows the same trend as that being observed for dc magnetic susceptibility measurements. There is no signature of the magnetic feature, at temperature $T^*$, below $T_{Eu}$. Newly these results are different from earlier reports that SC can emerge after suppressing SDW [10,22,28]. In these earlier reports, for TMA-flux-grown $Eu(Fe_{1-x}Co_x)_2As_2$ crystals, superconducting transition appears around 20 K although a zero-resistance is never reached. The lack of even zero resistance strongly suggests a very minor, second phase is responsible for the claimed SC. For tin-flux-grown $Eu(Fe_{1-x}Co_x)_2As_2$ crystals, zero resistance has been observed below 10 K in a narrow composition range of $0.15 \leq x \leq 0.20$ [11,25–28]. We do not see any SC across the range (specifically $x = 0.18$) and one only note that tin-flux-grown samples of other Co doped 122 arsenide have been fraught with problems [29].

Combining the results presented above and reported in the literature [6,11,25,26,36], the phase diagram of TMA-flux-grown $Eu(Fe_{1-x}Co_x)_2As_2$ single crystals is established. As shown in Fig. 7, here the SDW order of Fe is more quickly suppressed by Co doping. It only present up to Co-0.08. On the other hand, $T_{Eu}$ shows a systematic evolution with Co doping. At relatively low Co doping levels, it goes down and reaches a minimum at Co-0.08. For higher Co doping levels, it increases continuously up to Co-0.24. The substitution of Co for Fe in Eu-122 introduces electron carriers and contracts the lattice parameter $c$, which modifies both the indirect Ruderman–Kittel–Kasuya–Yosida (RKKY) interaction between the $Eu^{2+}$ moments

[37], as well as the direct interaction between the $Eu^{2+}$ and $Fe^{2+}$ moments. The changing tendency of $T_{Eu}$ with Co content is most probably due to the combined effects of these two changed interactions. A new magnetic feature appears at temperatures below $T_{Eu}$ and is gradually drawn to $T_{Eu}$ with Co doping higher than Co-0.10. A similar feature at alike temperature was shown for $Eu(Fe_{0.89}Co_{0.11})_2As_2$ crystals grown by TMA flux measured at applied field of 3 Oe parallel to *ab* plane, but not seen under same applied field parallel to *c* axis [38]. It was totally suppressed by the applied field of 1000 Oe parallel to *ab* plane. However there was no discussion of possible second magnetic transition in the text of Ref. 38. No signature of SC in the composition range from Co-0 to Co-0.24 has been observed. Therefore there is no coexistence of magnetism and SC in our case, which suggest that either the huge $Eu^{2+}$ moments work as the effective magnetic pair breaker, or that the $Eu^{2+}$ order totally suppresses necessary ingredients for the establishment of the superconducting state in $Eu(Fe_{1-x}Co_x)_2As_2$ grown by TMA flux. This is also consistent with our single-crystal $Sr_{1-y}Eu_y(Fe_{0.88}Co_{0.12})As_2$ work [24].

## IV. CONCLUSION

In summary, the crystal structure, magnetic, and electronic properties of $Eu(Fe_{1-x}Co_x)_2As_2$ single crystals grown by TMA flux have been carefully investigated. A phase diagram was established accordingly. With Co doping, the lattice parameter *c* shrinks linearly with Co doping. Meanwhile the SDW order of Fe is quickly suppressed, which only presents up to Co-0.08. It is found that $T_{Eu}$ shows a systematic evolution with Co doping, which first goes down and reaches a minimum at Co-0.08, then increases continuously up to Co-0.24. Such changing trend is probably due to the combined effects of RKKY interaction between the $Eu^{2+}$ moments and the interaction between Eu and $(Fe_{1-x}Co_x)As$ layers, which are both modified by the substitution of Co for Fe. Below $T_{Eu}$, a new magnetic feature is observed. We didn't observe any signature of SC in the composition range with $0 \leq x \leq 0.24$. We observed no coexistence of magnetism and SC.

**ACKNOWLEDGMENTS**

G. Wang would like to thank S. Manni, and U. S. Kaluarachchi for useful discussions and experimental assistance. This work was supported by the U.S. Department of Energy, Office of Basic Energy Science, Division of Materials Sciences and Engineering. The research was performed at the Ames Laboratory. Ames Laboratory is operated for the U.S. Department of Energy by Iowa State University under Contract No. DE-AC02-07CH11358. G. Wang was supported by the National Natural Science Foundation of China (Grant Nos. 51572291 and 51322211), the National Key Research and Development Program of China (Grant No. 2017YFA0302902), and the China Scholarship Council. WRM was supported by the Gordon and Betty Moore Foundations EPiQS Initiative through Grant GBMF4411.

∗ gangwang@iphy.ac.cn

[1] M. B. Maple, Phys. B, **215**, 110 (1995).

[2] P. C. Canfield, P. L. Gammel, and D. J. Bishop, Phys. Today, **51**, 40 (1998).

[3] M. R. Norman, Science, **332**, 196 (2011).

[4] D. N. Basov, and A. V. Chubukov, Nature Phys. **7**, 272 (2011).

[5] H. Raffius, E. Moersen, B. D. Morsel, W. Mueller-Warmuth, W. Jeitschko, L. Terbuechte, and T. Vomhof, J. Phys. Chem. Solids **54**, 135 (1994).

[6] Z. Ren, Z. W. Zhu, S. Jiang, X. F. Xu, Q. Tao, C. Wang, C. M. Feng, G. H. Cao, and Z. A. Xu, Phys. Rev. B **78**, 052501 (2008).

[7] Y. Xiao, Y. Su, M. Meven, R. Mittal, C. M. N. Kumar, T. Chatterji, S. Price, J. Persson, N. Kumar, S. K. Dhar, A. Thamizhavel, and Th. Brueckel, Phys. Rev. B **80**, 174424 (2009).

[8] Z. Ren, Q. Tao, S. Jiang, C. M. Feng, C. Wang, J. H. Dai, G. H. Cao, and Z. A. Xu, Phys. Rev. Lett. **102**, 137002 (2009).

[9] H. S. Jeevan, D. Kasinathan, H. Rosner, and P. Gegenwart, Phys. Rev. B **83**, 054511 (2011).

[10] S. Jiang, H. Xing, G. F. Xuan, Z. Ren, C. Wang, Z. A. Xu, and G. H. Cao, Phys. Rev. B **80**, 184514 (2009).

[11] M. Matusiak, Z. Bukowski, and J. Karpinski, Phys. Rev. B **83**, 224505 (2011).

[12] W. H. Jiao, Q. Tao, J. K. Bao, Y. L. Sun, C. M. Feng, Z. A. Xu, I. Nowik, I. Felner, and G. H. Cao, Europhys. Lett. **95**, 67007 (2011).

[13] W. H. Jiao, J. K. Bao, Q. Tao, H. Jiang, C. M. Feng, Z. A. Xu, and G. H. Cao, J. Phys.: Conf. Ser. **400**, 022038 (2012).

[14] U. B. Paramanik, D. Das, R Prasad, and Z Hossain, J. Phys.: Condens. Matter **25**, 265701 (2013).

[15] W. H. Jiao, H. F. Zhai, J. K. Bao, Y. K. Luo, Q. Tao, C. M. Feng, Z. A. Xu, and G. H. Cao, New J. Phys. **15**, 113002 (2013).

[16] H. S. Jeevan, Z. Hossain, D. Kasinathan, H. Rosner, C. Geibel, and P. Gegenwart, Phys. Rev. B **78**, 092406 (2008).

[17] Y. P. Qi, Z. S. Gao, L. Wang, D. L. Wang, X. P. Zhang, and Y. W. Ma, New J. Phys. **10**, 123003 (2008).

[18] C. F. Miclea, M. Nicklas, H. S. Jeevan, D. Kasinathan, Z. Hossain, H. Rosner, P. Gegenwart, C. Geibel, and F. Steglich Phys. Rev. B **79**, 212509 (2009).

[19] P N. Kurita, M. Kimata, K. Kodama, A. Harada, M. Tomita, H. S. Suzuki, T. Matsumoto, K. Murata, S. Uji, and T. Terashima, Phys. Rev. B **83**, 214513 (2011).

[20] K. Matsubayashi, K. Munakata, M. Isobe, N. Katayama, K. Ohgushi, Y. Ueda, N. Kawamura, M. Mizumaki, N. Ishimatsu, M. Hedo, I. Umehara, and Y. Uwatoko, Phys. Rev. B **84**, 024502 (2011).

[21] S. Zapf, and M. Dressel, Rep. Prog. Phys. **80**, 016501 (2017).

[22] J. J. Ying, T. Wu, Q. J. Zheng, Y. He, G. Wu, Q. J. Li, Y. J. Yan, Y. L. Xie, R. H. Liu, X. F. Wang, and X. H. Chen, Phys. Rev. B **81**, 052503 (2010).

[23] X. B. Chen, Z. A. Ren, H. S. Ding, and L. H. Liu, Sci. China **53**, 1212 (2010).

[24] R. W. Hu, S. L. Bud'ko, W. E. Straszheim, and P. C. Canfield, Phys. Rev. B **83**, 094520 (2011).

[25] A. Bachowski, K. Ruebenbauer, J. Zukrowski, Z. Bukowski, K. Rogacki, P. J. W. Moll, and J. Karpinski, Phys. Rev. B **84**, 174503 (2011).

[26] V. H. Tran, T. A. Zaleski, Z. Bukowski, L. M. Tran, and A. J. Zaleski, Phys. Rev.

B **85**, 052502 (2012).

[27] W. T. Jin, S. Nandi, Y. Xiao, Y. Su, O. Zaharko, Z. Guguchia, Z. Bukowski, S. Price, W. H. Jiao, G. H. Cao, and Th. Brückel, Phys. Rev. B **88**, 214516 (2013).

[28] W. T. Jin, Y. Xiao, Z. Bukowski, Y. Su, S. Nandi, A. P. Sazonov, M. Meven, O. Zaharko, S. Demirdis, K. Nemkovski, K. Schmalzl, Lan Maria Tran, Z. Guguchia, E. Feng, Z. Fu, and Th. Brückel, Phys. Rev. B **94**, 184513 (2016).

[29] S. Ran, S. L. Bud'ko, W. E. Straszheim, J. Soh, M. G. Kim, A. Kreyssig, A. I. Goldman, and P. C. Canfield, Phys. Rev. B **85**, 224528 (2012).

[30] N. Ni, M. E. Tillman, J.-Q. Yan, A. Kracher, S. T. Hannahs, S. L. Bud'ko, and P. C. Canfield, Phys. Rev. B **78**, 214515 (2008).

[31] S. Ran, Ph.D. thesis, Iowa State University, 2014.

[32] P. C. Canfield, T. Kong, U. S. Kaluarachchi, and N. H. Jo, Philos. Mag. 96, **84** (2016).

[33] A. Jesche, M. Fix, A. Kreyssig, W. R. Meier, and P. C. Canfield, Philos. Mag. **96**, 2115 (2016).

[34] Y. Su, P. Link, A. Schneidewind, Th. Wolf, P. Adelmann, Y. Xiao, M. Meven, R. Mittal, M. Rotter, D. Johrendt, Th. Brueckel, and M. Loewenhaupt, Phys. Rev. B **79**, 064504 (2009).

[35] H. S. Jeevan, Z. Hossain, D. Kasinathan, H. Rosner, C. Geibel, and P. Gegenwart, Phys. Rev. B **78**, 052502 (2008).

[36] Z. Guguchia, J. Roos, A. Shengelaya, S. Katrych, Z. Bukowski, S. Weyeneth, F. Muranyi, S. Strassle, A. Maisuradze, J. Karpinski, and H. Keller, Phys. Rev. B **83**,

144516 (2011).

[37] A. Akbari, P. Thalmeier, and I. Eremin, New J. Phys. **15**, 033034 (2013).

[38] I. Nowik, I. Felner, Z. Ren, G. H. Cao, and Z. A. Xu, New J. Phys. **13**, 023033 (2011).

TABLE I. The WDS analyses of Co content for $Eu(Fe_{1-x}Co_x)_2As_2$ crystals. $\sigma$ is the standard deviation of the 12 measured values.

| Nominal Co content ($x$) | 0 | 0.04 | 0.08 | 0.10 | 0.12 | 0.18 | 0.24 |
|---|---|---|---|---|---|---|---|
| Co/(Fe+Co) by WDS analyses | 0 | 0.041 | 0.074 | 0.090 | 0.117 | 0.174 | 0.235 |
| $\sigma$ | 0 | 0.001 | 0.001 | 0.001 | 0.001 | 0.001 | 0.002 |

FIG. 1 Photograph of Eu(Fe$_{1-x}$Co$_x$)$_2$As$_2$ single crystal with x = 0.18 (Co-0.18). The scale bar is 1 mm.

FIG. 2 (a) The XRD patterns of Eu(Fe$_{1-x}$Co$_x$)$_2$As$_2$ crystals showing (*00l*) diffraction peaks. The intensity is on log scale and the spectra are offset for clarity. (b) The enlarged (*008*) diffraction peak, the shoulder is due to the Cu $K_{\alpha 2}$ radiation.

FIG. 3 The Co-doping-dependent lattice parameter *c* for Eu(Fe$_{1-x}$Co$_x$)$_2$As$_2$ crystals grown by TMA flux [6] and tin flux [25,26]. The solid line is the linear fit to the data of this work.

FIG. 4 (a) The temperature-dependent dc magnetic susceptibility of Eu(Fe$_{1-x}$Co$_x$)$_2$As$_2$ single crystals below 25 K measured under a applied field of 10 Oe parallel to the *ab* plane in the ZFC protocol. (b) The *d(χT)/dT* curve for Co-0.24. The arrows denote the magnetic transition temperatures, which are determined by minimum *d(χT)/dT*.

FIG. 5. The field-dependent magnetization for Eu(Fe$_{1-x}$Co$_x$)$_2$As$_2$ single crystals at 5 K measured under applied field parallel to the *ab* plane. The dotted line shows the theoretical saturated magnetic moment of Eu$^{2+}$ (7.0 $\mu_B$).

FIG. 6. (a) The temperature-dependent normalized in-plane resistance $R_{ab}$ for Eu(Fe$_{1-x}$Co$_x$)$_2$As$_2$ single crystals. (b) The temperature derivative of resistance for Co-0. $T_{SDW}$ and $T_{Eu}$ are determined by *dR/dT*, denoting the temperatures for SDW order and the ordering of Eu$^{2+}$ moments.

FIG. 7. The phase diagram of TMA-flux grown Eu(Fe$_{1-x}$Co$_x$)$_2$As$_2$ single crystals. The transition temperatures (open symbols) for TMA-flux-grown and tin-flux-grown Eu(Fe$_{1-x}$Co$_x$)$_2$As$_2$ crystals from literatures are shown for comparison [6,11,25,26,36].

$T_{SDW}$, $T_{Eu}$, and $T_c$ denote the SDW order temperature of Fe, ordering temperature of Eu$^{2+}$ moments, and superconducting transition temperature, respectively. $T^*$ denotes the magnetic transition at temperatures below $T_{Eu}$. The vertical axis is in log scale. The line is just a guide to the eye.

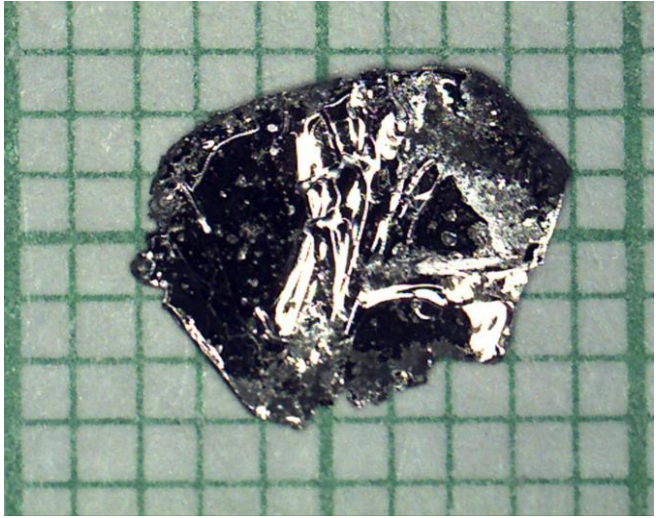

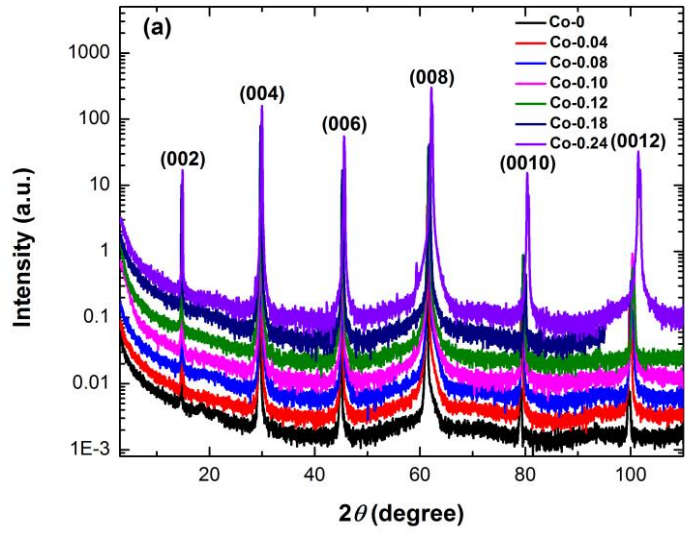

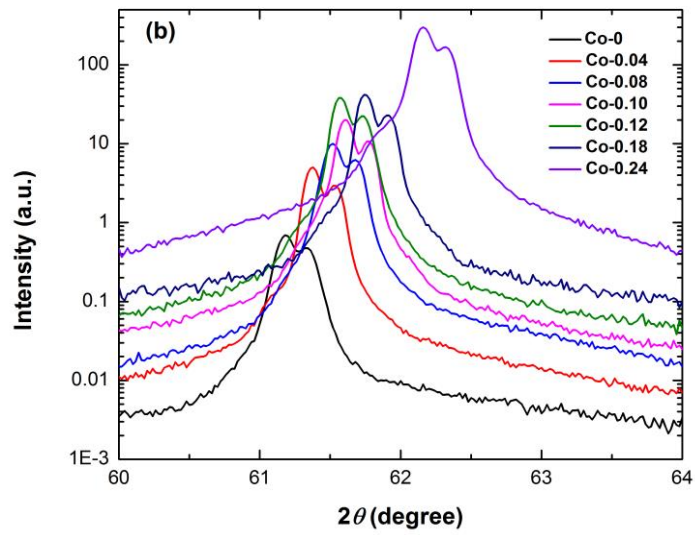

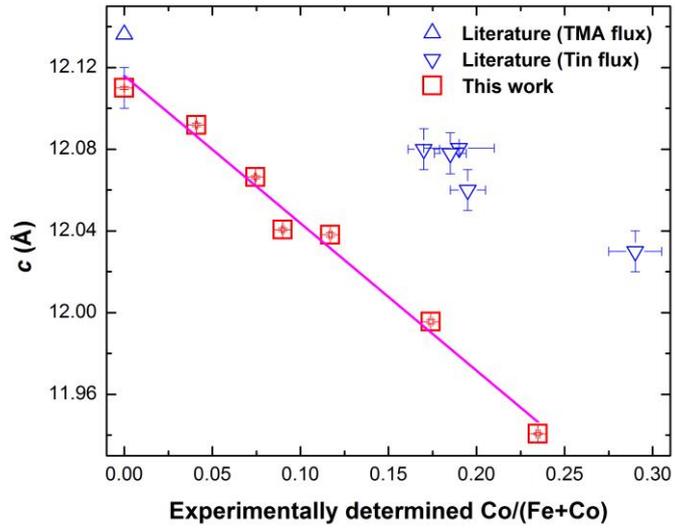

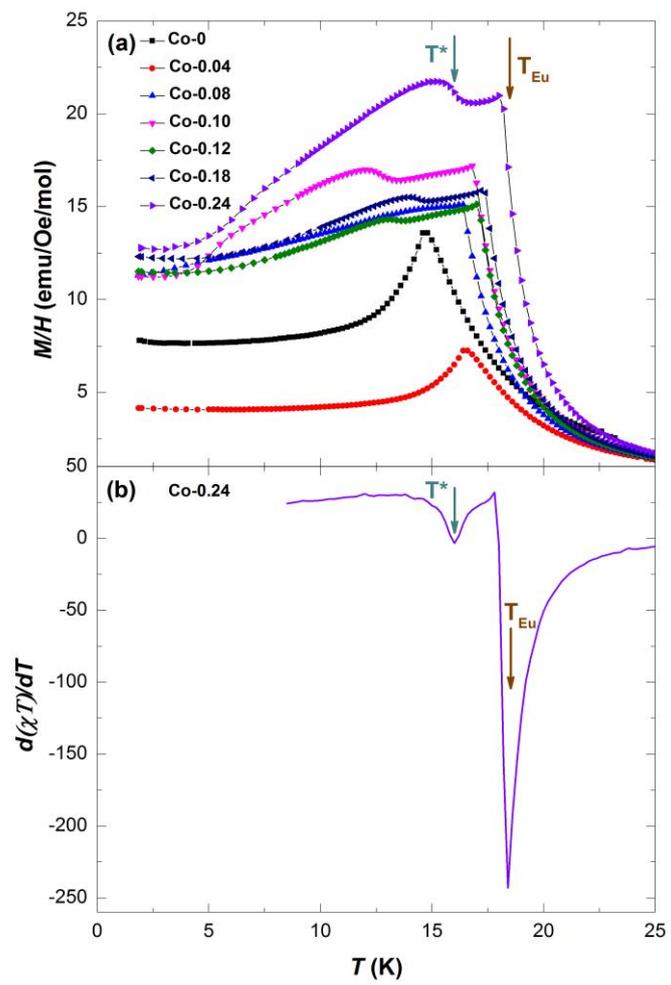

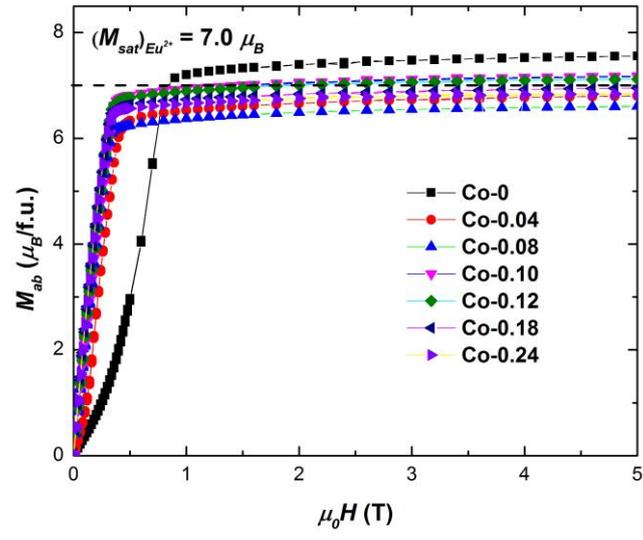

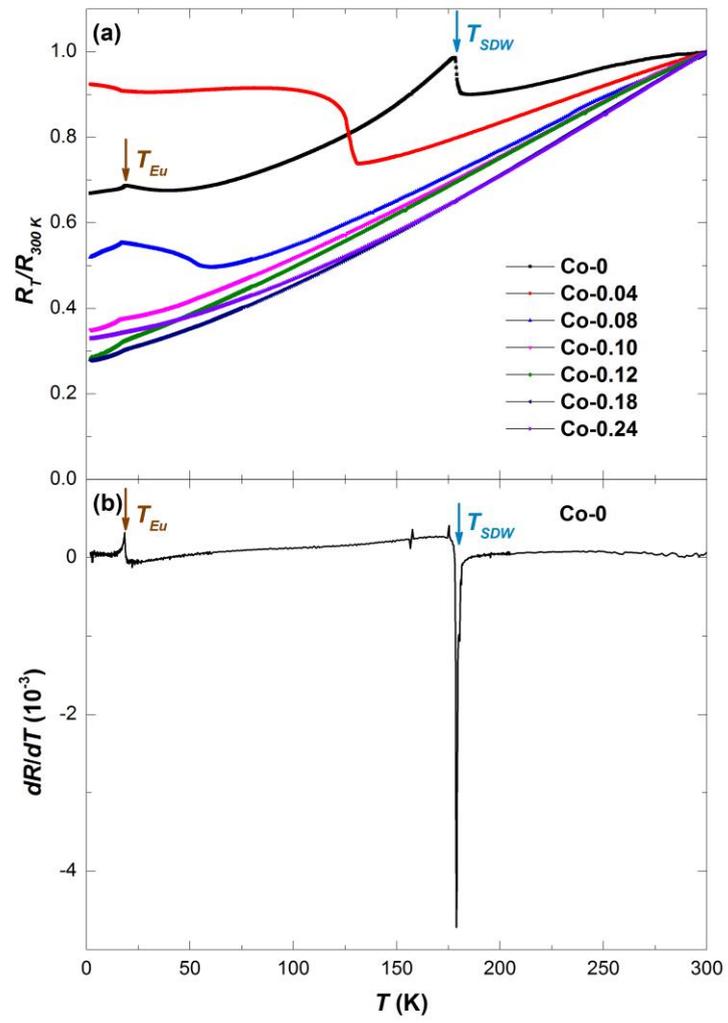

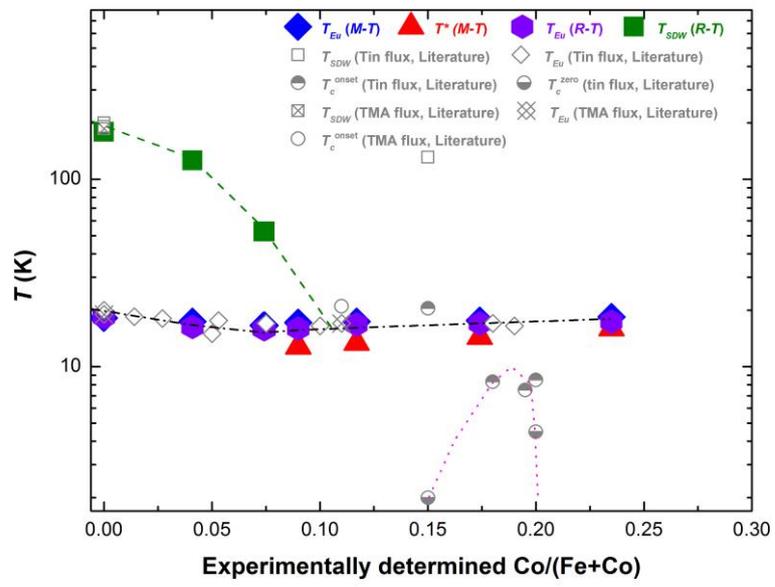